\documentclass[aps,prd,nofootinbib,twocolumn]{revtex4}
\usepackage{amssymb}
\usepackage{amsmath,color}
\usepackage{epsfig}
\usepackage{graphicx,epsf,epsfig}
\usepackage{bbm}
\usepackage{subfigure}
\usepackage{multirow}
\usepackage{setspace}
\usepackage{verbatim}
\usepackage{epstopdf}
\usepackage[linktocpage]{hyperref}
\usepackage{multirow}
\newcommand{\be}{\begin{eqnarray}}
\newcommand{\ee}{\end{eqnarray}}
\newcommand{\bq}{\begin{equation}}
\newcommand{\eq}{\end{equation}}
\newcommand{\bp}{\begin{split}}
\newcommand{\ep}{\end{split}}
\newcommand{\mrm}{\mathrm}

\begin{document}

\title{Primordial Black Holes from Cosmic Domain Walls}

\author{Jing Liu$^{1,2}$}
\email{liujing@itp.ac.cn}

\author{Zong-Kuan Guo$^{1,2}$}
\email{guozk@itp.ac.cn}

\author{Rong-Gen Cai$^{1,2}$}
\email{cairg@itp.ac.cn}

\affiliation{$^1$CAS Key Laboratory of Theoretical Physics, Institute of Theoretical Physics,
	Chinese Academy of Sciences, P.O. Box 2735, Beijing 100190, China}
\affiliation{$^2$School of Physical Sciences, University of Chinese Academy of Sciences,
	No.19A Yuquan Road, Beijing 100049, China}

\begin{abstract}
	We investigate the formation of primordial black holes (PBHs) from the collapse of spherically symmetric domain wall bubbles, which spontaneously nucleate via quantum tunneling during inflation.
	Since the tension of domain walls changes with time and so domain walls nucleate in a short time interval,
	the mass function of PBHs in general has a spike-like structure.
	In contrast to models in which PBHs produced from overdense regions, our model avoids the uncertainties of PBHs production mechanism. PBHs from domain walls with mass around $10^{20}\mrm{g}$ may constitute all dark matter, those with mass around $10^{34}\mrm{g}$ can explain the merger events of binary black holes detected by LIGO.
\end{abstract}

\maketitle

\section{Introduction}
\label{sec:int}

Dark matter is a key ingredient in the standard model of modern cosmology. However,  the nature of dark matter  still remains mysterious.
The primordial black hole  (PBH) formed in the early universe is an interesting candidate of dark matter.  It is also an economical solution to the dark matter 
problem  without new physics if the PBHs constitute all dark matter.   Recently it has been shown that PBHs can explain the merger events of binary black holes observed by  LIGO (see Refs.~\cite{Sasaki:2018dmp,Carr:2016drx} for recent reviews).

In the early universe, PBHs may form due to the gravitational collapse of large enough density fluctuations in some overdense regions. This happens when the large curvature fluctuations at small scales produced during inflation, reenter the Hubble horizon after inflation. The large curvature fluctuations at scales much smaller than the CMB scale can be 
produced in various models, for instance, a period of ultra-slow-roll~\cite{Gao:2018pvq,Xu:2019bdp,Gong:2017qlj,Addazi:2018pbg,Dalianis:2018frf,Hertzberg:2017dkh,Cicoli:2018asa,Fu:2019ttf}, sound speed resonance~\cite{Cai:2018tuh} and an early dust-like stage~\cite{Cotner:2018vug,Khlopov:2008qy} etc.. The amplification of density fluctuations can also be caused by subhorizon nonlinear dynamics such as phase transition~\cite{Jedamzik:1999am,Rubin:2000dq,Upadhyay:1999vk,Cai:2017tmh, Cai:2017cbj} and preheating~\cite{Bassett:2000ha,Green:2000he,Liu:2017hua,Liu:2018rrt}. However, the mechanism of PBH formation from overdense regions has some uncertainties.
It is known that the abundance of PBHs from the gravitational collapse is extremely sensitive to the threshold which can be  fixed by numerical simulations~\cite{Carr:1975qj,Harada:2013epa,Carr:2016drx,Niemeyer:1997mt,Niemeyer:1999ak,Green:2004wb,Musco:2004ak}. In addition, 
recently it has been pointed out that, for a given amplitude of the power spectrum, the abundance of PBHs generated from a narrow peak is exponentially smaller than that from a board peak~\cite{Germani:2018jgr}.
The choice of window function and time at which perturbations are evaluated can also affect the abundance of PBHs~\cite{Young:2019osy}.

PBHs can also be produced from cosmic domain walls (DWs). The formation of black holes and wormholes from vacuum bubbles of DWs is studied in Refs.~\cite{Garriga:2012bc,Garriga:2015fdk,Deng:2016vzb} and a numerical simulation is made  in Ref.~\cite{Deng:2016vzb}.
In Ref.~\cite{Ge:2019ihf} the author studies the formation of PBHs from the collapse of closed DWs in QCD axion models. DWs are sheet-like objects in three spatial dimensions 
and can be formed when a discrete symmetry is spontaneously broken. Some new physics models beyond the standard model of particle physics allow the existence of DWs since a discrete symmetry is pervasive in high-energy physics. The scalar field settles in different vacua on each side of a DW.
Since the energy stored in DWs is proportional to their area, the stable DWs will eventually dominate the universe which leads to a serious cosmological problem~\cite{Zeldovich:1974uw}.
To avoid this, one considers the production of PBHs from the spherical DW bubbles which nucleate during inflation.  When the Hubble radius exceeds the DW radius, they become unstable and form PBHs. Since the nucleation rate can be exponentially suppressed by its Euclidean action, it is possible that DWs become unstable before they overclose the universe.

The mass function of PBHs from spherical DW bubbles is given in Ref.~\cite{Garriga:2015fdk} where the tension of DWs is a constant during inflation.
Such a mass function is very broad, therefore the abundance of PBHs could be strictly constrained by observations.
In this paper, we propose a two-field inflationary model in which the tension of DWs changes with time during inflation.
Since the DW nucleation happens in a short period of time when the Euclidean action of DWs reaches minimum, the mass function has a spike-like structure. We find that in our model PBHs could constitute all dark matter for some constraint windows.

It is known that the gravitational waves seeded by scalar fluctuations at second order, which will collapse to form PBHs when they reenter the horizon,  during inflation can be used to constrain the abundance of PBHs~\cite{Cai:2018dig,Saito:2008jc,Saito:2009jt,Wang:2019kaf,Ananda:2006af,Baumann:2007zm,Cai:2019elf,Yuan:2019udt,Lu:2019sti}.
On the other hand,  let us notice that according to Birkhoff's theorem, PBHs produced by spherical DWs collapse will not  produce stochastic gravitational wave background. In this way, the abundance of PBHs could avoid the constraint from the gravitational wave observation constraint.


The paper is organized as follows. In Sec.~\ref{sec:PBH}, we present the mechanism for the PBH formation from cosmic DWs.
In Sec.~\ref{sec:inf}, we propose a two-field inflationary model to produce DWs in a short period of time. In Sec.~\ref{sec:mas}, we calculate the mass function and give the parameter space in which PBHs may constitute all dark matter. In Sec.~\ref{sec:con}, we summarize our results.

\section{PBHs from Cosmic DWs}
\label{sec:PBH}
In this section, we present the theoretical formulas of cosmic DWs and the dynamics of the PBH formation from cosmic DWs in the radiation-dominated (RD) era and the matter-dominated (MD) era. In the following we shall adopt the notation of Ref.~\cite{Deng:2016vzb}.

DWs are two-dimensional topological defects which may form in the early universe
when a discrete symmetry is spontaneously broken. Consider a simple model with a real scalar field $ \phi $ as an illustration
\begin{eqnarray}
\mathcal{L}=-\frac{1}{2} \partial^{\mu} \phi \partial_{\mu} \phi-V(\phi),\nonumber\\
V(\phi)=\frac{\lambda}{4}\left(\phi^{2}-v^{2}\right)^{2}.
\end{eqnarray}
Such a potential has two degenerate minima when the $Z_{2}$ symmetry is spontaneously broken.
On each side of the DW the field $\phi$ takes different vacuum expectation values, i.e., $\phi=\pm\nu$. The static DW solution vertical to the $x$-axis can be expressed as
\be
\phi(x)=v \tanh \left[\sqrt{\frac{\lambda}{2}} v x\right].
\ee
The field $\phi$ deviates from the vacuum expectation value and changes rapidly near $x=0$. The surface energy density of the DWs is
\be
\sigma=\frac{4}{3} \sqrt{\frac{\lambda}{2}} v^{3},
\ee
which is also referred to as the tension of DWs.

In the MD or RD era, if the interaction between $\phi$ and the background fluid is negligible, numerical results confirm the energy density of DWs scales as $\rho_{DW}\propto 1/t$~\cite{Leite:2011sc,Leite:2012vn,Martins:2016ois} and the energy density fraction of DWs scales as $\Omega_{DW}\propto t$. Once DWs dominate the universe, the Hubble horizon begins to shrink which conflicts with present observable. To get rid of the problem, we consider spherically symmetric DW bubbles spontaneously nucleated during inflation. The number density of spherical DWs is exponentially suppressed by its Euclidean action, so that they collapse to PBHs before becoming dominant in the whole universe.
Applying the thin wall approximation, the metric of the corresponding planar DW is given by~\cite{Vilenkin:1984hy,Ipser:1983db}
\begin{eqnarray}
d s^{2}=&&-\left(1-\frac{|x|}{t_{\sigma}}\right)^{2} d t^{2}+d x^{2}\nonumber\\
&&+\left(1-\frac{|x|}{t_{\sigma}}\right)^{2} e^{2 t / t_{\sigma}}\left(d y^{2}+d z^{2}\right),
\end{eqnarray}
where
\begin{equation}
t_{\sigma}=\dfrac{1}{2\pi G\sigma}.
\end{equation}

In the case of $t\ll t_{\sigma}$, the energy density fraction of DWs $\Omega_{DW}\ll 1$ and the metric of surrounding universe is unaffected by DWs. In the opposite case, the surrounding universe is dominated by DWs and the metric is determined by DWs. To estimate whether DWs are dominant at the time of the gravitational collapse, we define a time $t_{H}$ when the Hubble radius exceeds the radius of DWs. The condition $t_{H}<t_{\sigma}$ is denoted as the subcritical case while the case $t_{H}>t_{\sigma}$ is denoted as the supercritical one.

In the subcritical case, heuristically speaking, since the potential energy is stored in DWs,
the potential energy is transferred to the kinetic energy, the DW bubbles will shrink. The initial mass of PBHs, $M_{i}$, has the same order of the total mass of the DW bubble $M_{i}=4 \pi \sigma C R^{2}(t_{H})$, where $C=0.62$ in the RD era and $C=0.15$ in the MD era~\cite{Garriga:2015fdk}. The PBH mass evolves with time due to accretion once they are formed. According to numerical results of Ref.~\cite{Deng:2016vzb}, in the RD era the final mass simply doubles $M_{f,RD}\approx 2M_{i,RD}$. In this case, the PBH mass is independent of the formation time. Since $\Omega_{DW}\ll 1$, the PBH mass is smaller than the Hubble mass. In the MD era, it is shown that the matter initially surrounded by DWs will finally be absorbed by PBHs and $M_{f,MD}=4\pi R^{3}(t_{e})H^{2}(t_{e})M_{\mrm{p}}^{2}$, where $ R(t_{e}) $ and $ H(t_{e}) $ are the physical radius of DWs and the Hubble parameter at the end of inflation, respectively.

In the supercritical case, since DW perturbations propagates outwards through rarefaction (or decompression) wave, PBHs form when the Hubble horizon exceeds the wave front. The horizon of the initial black hole cannot exceed the Hubble radius at the formation time, this gives an upper bound on the initial PBH mass. The numerical results indicate that the actual mass is close to the upper bound given in~\cite{Deng:2016vzb}. In the MD era, the speed of wave is much smaller than one, so the upper bound of the PBH horizon is the DW radius at the PBH formation time. Thus, the upper bound of the initial mass is $M_{i,MD}\approx 4\pi R^{3}(t_{e})H(t_{e})^{2}M_{\mrm{p}}^{2}$. DWs blow the matter away and build up an empty layer. As the PBH is formed, the accretion is interdicted by the empty layer and the mass remains constant, $M_{f,MD}\approx M_{i,MD}$. In the RD era, the speed of wave is the speed of sound. For large $R(t_{e})$, $M_{i,RD}\approx2.8\times 8\pi R^{2}(t_{e})H(t_{e})M_{\mrm{p}}^{2}$ which is confirmed by numerical results in Ref.~\cite{Deng:2016vzb}. The final mass is $M_{f,RD}\approx2M_{i,RD}$.

\section{Model Building}
\label{sec:inf}
We consider a two-field inflationary model in which the inflaton $\phi$ is minimally coupled to gravity
\begin{eqnarray}
S=\int d^{4} x \sqrt{-g} &&\left[ -\frac{M_{\mathrm{p}}^{2}}{2} R+\frac{1}{2} \partial_{\mu} \phi \partial^{\mu} \phi+\frac{1}{2} \partial_{\mu} \chi \partial^{\mu} \chi \right.\nonumber\\
&&\left. +V(\phi,\chi)\right],
\end{eqnarray}
where the effective potential $V(\phi,\chi)$ reads

\begin{equation}
V(\phi,\chi)=\dfrac{\lambda_{\chi}}{4}\left[\chi^{2}-\alpha^{2}\left(\phi-\phi_{c}\right)^{2}-m^{2} \right]^{2}+f(\phi).
\end{equation}
The last term is an effective potential of the inflaton $\phi$ which should be compatible with the Planck results~\cite{Akrami:2018odb}.

\begin{figure}[t]
	\includegraphics[width=3in]{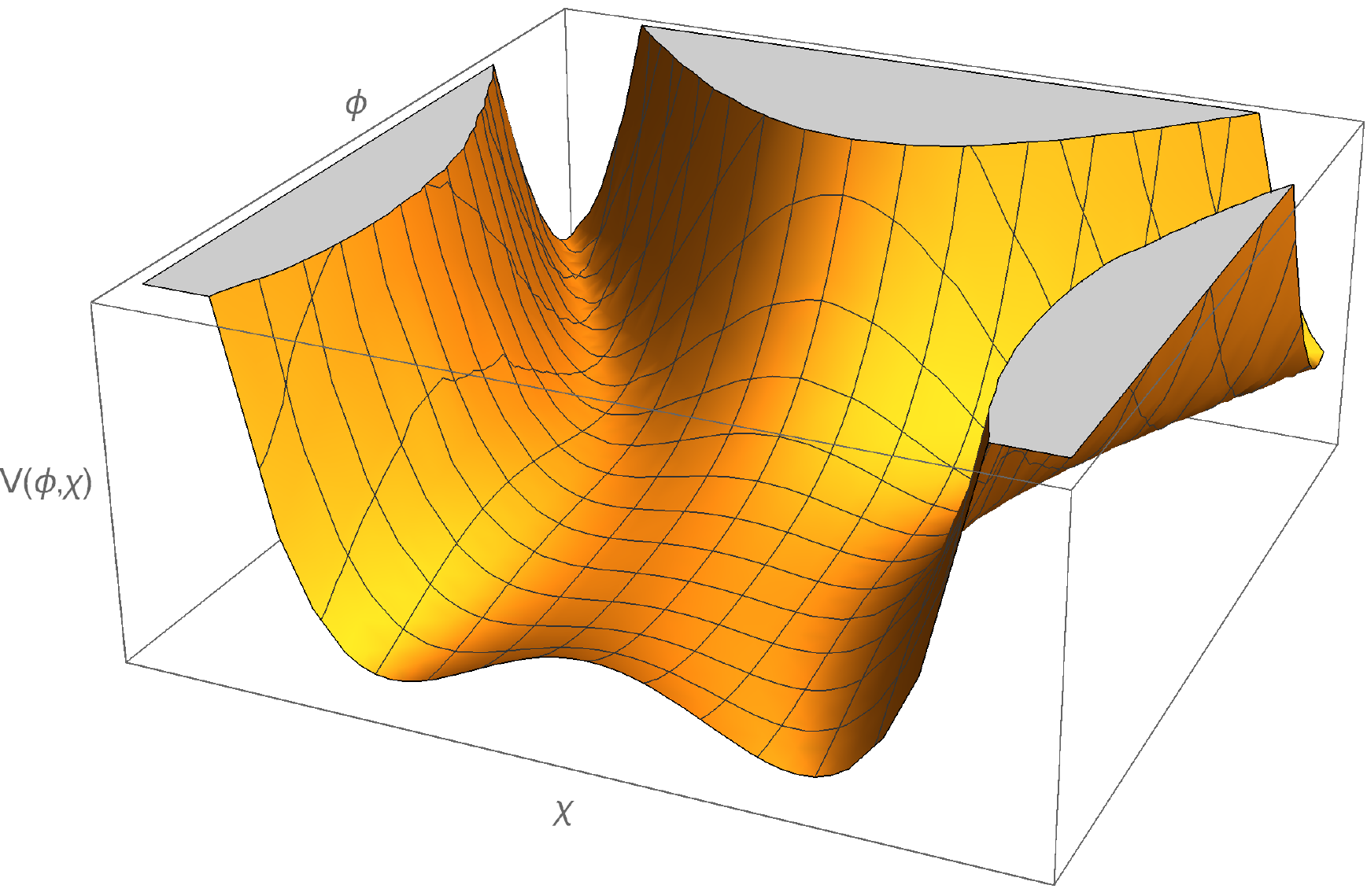}
	\caption{Potential $V(\phi,\chi) $.}
	\label{fig:V}
\end{figure}

The effective potential $V(\phi,\chi)$ provides two degenerate vacua in the $ \chi $ direction as shown in Fig.~\ref{fig:V}. Initially $\chi$ takes one of the minima. During inflation, the dynamics of the inflaton $ \phi $ remains unaffected by $ \chi $. Neglecting the spatial gradient of the scalar fields during inflation, the Friedman equation and the equations of motion of the scalar fields are
\begin{eqnarray}
&&H^{2}=\nonumber
\frac{1}{3 M_{\mathrm{p}}^{2}}\left(\frac{1}{2} \dot{\phi}^{2}+ \frac12 \dot{\chi}^{2}+V(\phi,\chi)\right),\nonumber\\
&&\ddot{\phi}+3 H \dot{\phi}+\frac{\partial V}{\partial \phi}=0,\nonumber\\
&&\ddot{\chi}+3 H \dot{\chi}+\frac{\partial V}{\partial \chi}=0.
\end{eqnarray}
We choose $\lambda_{\chi}=0.3$ and $\alpha=3\times10^{-5}$ in the following.
We consider three parameter sets of $ \phi_{c} $ and $ m $ listed in Table~\ref{ta:para} as examples.
\begin{table}
	\begin{tabular}{|c|c|c|}
	\hline
	  Set&$\phi_{c}/M_{\mathrm{p}}$& $m/M_{\mathrm{p}}$   \\
	\hline
	  $1$ &$3.74$ & $3.24\times10^{-5}$   \\
	\hline
	  $2$ &$4.17$ & $3.16\times10^{-5}$   \\
	\hline
	  $3$ &$5.50$ & $2.98\times10^{-5}$   \\
	\hline
	\end{tabular}
\caption{The parameter sets we choose as examples. }
\label{ta:para}
\end{table}

Consider a power-law inflation
\begin{equation}
f(\phi)=\dfrac{\lambda_{\phi}}{p}\phi^{p},
\end{equation}
with $p=2/5$~\cite{Silverstein:2008sg}. The initial value of the inflaton is set to be $ \phi_{i}=6.25M_{\mrm{p}} $ and inflation ends at $ \phi_{e}=M_{\mrm{p}} $ so that the number of $e$-folds $ N=50 $. The predicted scalar spectral index $ n_{s}=0.976 $ and tensor-to-scalar ratio $ r=0.03 $ are in agreement with the CMB data~\cite{Akrami:2018odb}.

\begin{figure}[h!]
	\includegraphics[width=3.2in]{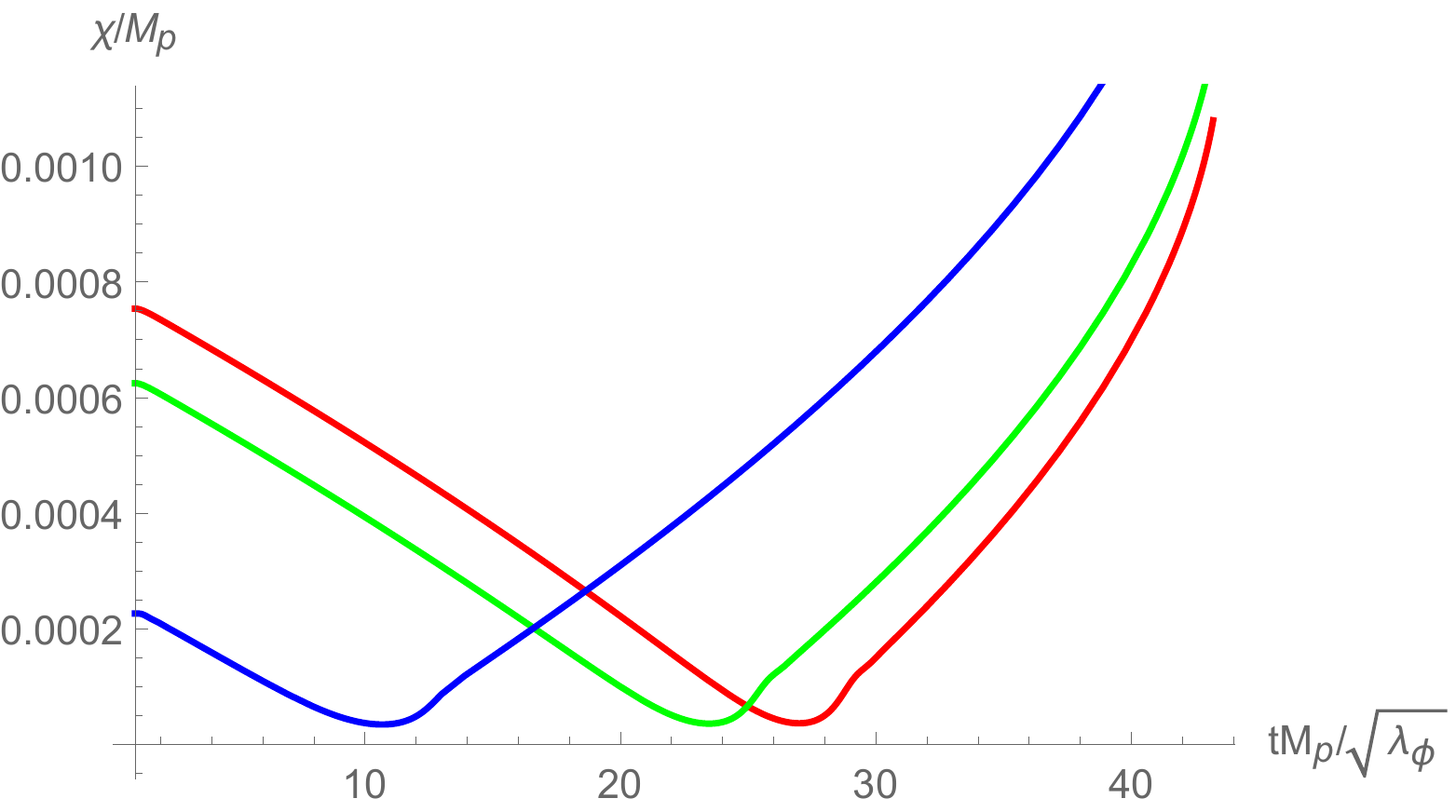}
	\caption{Evolutions of $ \chi $ for the parameter set $ 1 $ (red), set $ 2 $ (green), set $ 3 $ (blue) in Table~\ref{ta:para}.}
	\label{fig:chi}
\end{figure}

For the three parameter sets, the evolutions of $ \chi $ is shown in Fig.~\ref{fig:chi}.
The field $ \chi $ lies in one of the the minima of the potential due to its large mass. The vacuum expectation value of $\chi$ depends on $\phi(t)$, which reaches the minimum at the time when $\phi=\phi_{c}$. The large mass of $ \chi $ in the beginning of inflation makes the model free from CMB-scale constrains of non-Gaussianity and entropy perturbations.

We can check that the adiabatic approximation is valid around the time when the mass of $\chi$ reaches the minimum. This can be manifested by estimating
\be
\dfrac{\dot{m}_{\chi}}{m_{\chi}^{2}}\approx \dfrac{\alpha^{2}\left(\phi-\phi_{c}\right)\dot{\phi}}{m_{\chi}^{3}}.
\ee
Using typical energy scale of inflation $ 10^{-4}M_{\mrm{p}}$ and the spectrum of scalar perturbations $\mathcal{P}_{\mathcal{R}}\approx 2\times 10^{-9}$, it is obtained by
\be
\dot{\phi}\sim 10^{-12}M_{\mrm{p}}^{2}.
\ee
The adiabatic approximation can be examined at the time when $m=\alpha(\phi-\phi_{c})$,
\be
\dfrac{\dot{m}_{\chi}}{m_{\chi}^{2}}\approx 10^{-12}\dfrac{\alpha}{m^{2}}\ll 1.
\ee
Therefore, the adiabatic approximation is valid during inflation.

\section{Mass function of PBHs}
\label{sec:mas}
In this section, we derive the mass function of PBHs and compare it to the observational constrains.
Quantum nucleation of topological defects during inflation was studied in Ref.~\cite{Basu:1991ig}, where the authors derived the nucleation rate of cosmic strings and cosmic DWs in a de Sitter background spacetime.
The Euclideanized de Sitter space is a four-sphere of radius $H^{-1}$, and DWs nucleated during inflation by quantum tunneling can be described as a three-sphere with the maximal radius, i.e., $H^{-1}$. The Euclidean action is proportional to the surface area
\begin{equation}
S_{E}(t)=2\pi^{2}\sigma(t) H^{-3}(t).
\label{eq:S_E}
\end{equation}
The nucleation rate per unit physical volume per unit time can be written as
\begin{equation}
\lambda(t)=H^{4}(t)Ae^{-S_{E}(t)},
\label{eq:lambda}
\end{equation}
which is valid as long as the DWs can be treated semiclassicaly, i.e., $\sigma\gtrsim H^{3}$. The nucleation rate is suppressed exponentially if $\sigma\gg H^{3}$. In the opposite case, the DW bubbles is overproduced and finally dominate the universe. Therefore, we are interested in the case where $\sigma$ and $ H^{-3}$ are of the same order. Here $A$ is a slowly
varying function of $\sigma H^{-3}$ and one expects $A\sim 1$~\cite{Garriga:1993fh,Basu:1991ig}. Then the number density of DWs is
\begin{equation}
dN=\lambda(t_{*})a^{3}(t_{*})d^{3}\mathbf{x}dt_{*},
\end{equation}
where $t_{*}$ denotes the nucleation time.

\begin{figure}[h!]
	\includegraphics[width=3.2in]{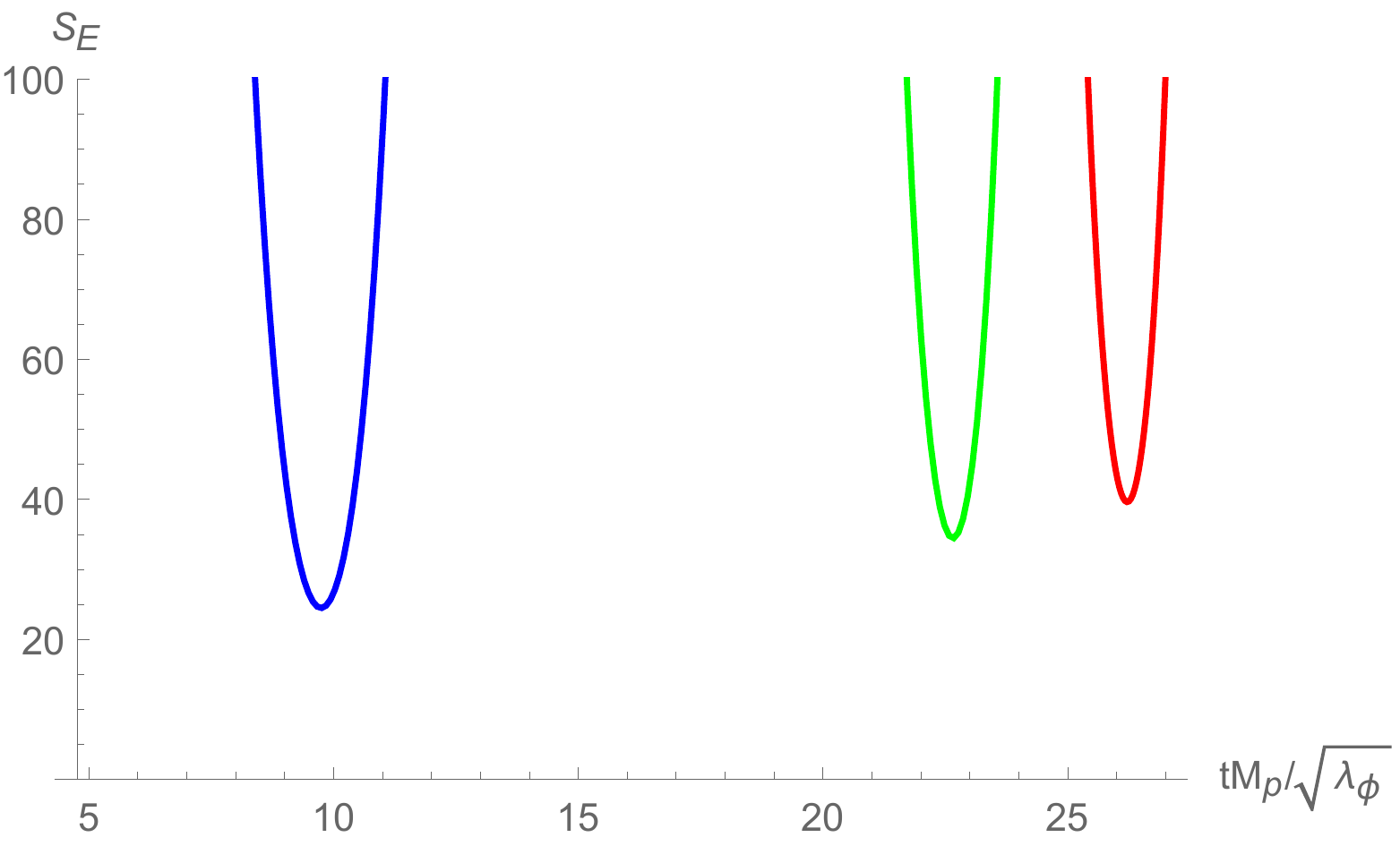}
	\caption{Evolutions of $ S_{E} $ for the parameter set $ 1 $ (red), set $ 2 $ (green), set $ 3 $ (blue) in Table~\ref{ta:para}.}
	\label{fig:S_E}
\end{figure}

The evolutions of the Euclidean action are shown in Fig.~\ref{fig:S_E}. At the time $\phi=\phi_{c}$, $ S_{E} $ reaches the minimum, which depends on $ m $. During inflation $ S_{E} $ is larger than one, so the semiclassical approximation is valid.

DWs nucleated at the end of inflation has the smallest radius, $1/H(t_{e})$, which collapse into PBHs with smallest mass shortly after inflation ends.
Following the results in Sec.~\ref{sec:PBH}, the minimum mass is
\be
M_{min}=4\pi \sigma CH(t_{e})^{-2}.
\ee
In our three parameter sets, $\sigma(t_{e})\sim 10^{-5} M_{\mathrm{p}}^{3}$, and $M_{min}\sim 10^{-8}\mrm{g}$.
The upper bound of PBH mass in the subcritical case is
\be
M_{max}=4\pi C \sigma(t_{e}) H^{-2}(t_{\sigma})\approx 10^{2}\mrm{g}.
\ee
PBHs with smaller mass than $10^{15}\mrm{g}$ have been already completely evaporated due to Hawking radiation.
For the mass range of our interest, therefore we focus on the supercritical case.

Following the results in Sec.~\ref{sec:PBH}, the final mass of PBHs from DWs is approximately
\begin{equation}
M=5.6\times 8\pi R^{2}(t_{e})H(t_{e})M_{\mrm{p}}^{2}.
\label{eq:mass}
\end{equation}
Since the universe is radiation-dominated from the end of inflation to the matter-radiation equality, the Hubble mass increases with time. In the supercritical case, the PBH mass approximately equals to the Hubble mass, so it increases with the PBH formation time. The mass of PBHs formed at matter-radiation equality is estimated by $M_{\mrm{eq}}\sim 10^{50}\mathrm{g}$ which is many orders of magnitude larger than the supermassive black holes and is strongly disfavored by CMB data. This upper bound implies the PBHs from DWs can provide seeds for super massive black holes~\cite{Duechting:2004dk}. In the following we only focus on PBHs formed in the RD era.


\begin{figure}[h!]
	\includegraphics[width=3.2in]{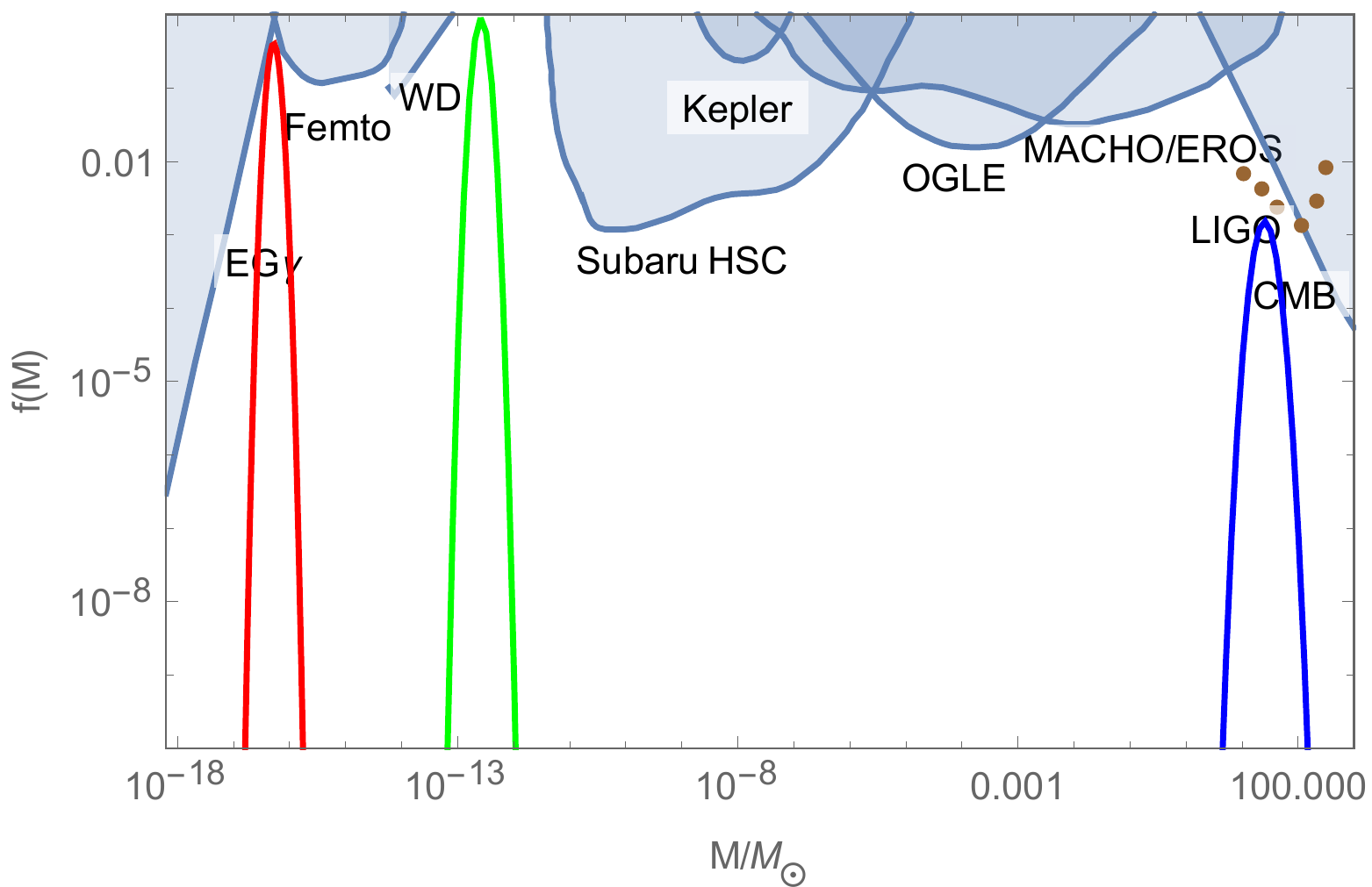}
	\caption{The mas functions of PBHs for the parameter set $ 1 $ (red), set $ 2 $ (green), set $ 3 $ (blue) in Table~\ref{ta:para}. The current constraints include the extra-galactic gamma-rays from the PBH evaporation (EG$ \gamma $)~\cite{Carr:2009jm}, the femtolensing of gamma-ray bursts(Femto)~\cite{Barnacka:2012bm}, the white dwarfs explosion (WD)~\cite{Graham:2015apa}, the microlensing events with Subaru HSC (Subaru HSC)~\cite{Niikura:2017zjd}, with Kepler satellite (Kepler)~\cite{Griest:2013esa}, with MACHO/EROS~\cite{Allsman:2000kg,Tisserand:2006zx}, with OGLE~\cite{Niikura:2019kqi}, the CMB measurements~\cite{Poulin:2017bwe,Ali-Haimoud:2016mbv}, and the LIGO merger rate set by LIGO O1~\cite{Ali-Haimoud:2017rtz}.}
	\label{fig:mass_func}
\end{figure}

We use the mass function $f(M)$ to characterize the fraction of PBHs against the whole dark matter at present per logarithmic mass interval
\be
f(M)\equiv \dfrac{1}{\rho_{\mrm{DM}}}\dfrac{d\rho_{\mrm{PBH}}(M)}{d\ln M}=\dfrac{1}{\rho_{\mrm{DM}}}\dfrac{M^{2}dN}{a^{3}(t_{0})d^{3}\mathbf{x}dM},
\ee
where $ t_{0} $ is the present time.

At the end of inflation, the physical radius of DWs nucleated at the time $t_{*}$ is
\be
\label{eq:Rte}
R(t_{e})=\dfrac{1}{H(t_{*})}\dfrac{a(t_{e})}{a(t_{*})}.
\ee
In the standard slow-roll inflationary scenario, the Hubble parameter changes slightly during inflation. The radius of DWs at the nucleation time is almost the same, thus $R$ is larger for the DWs formed earlier according to Eq.~\eqref{eq:mass}.
Substituting Eq.~\eqref{eq:Rte} into Eq.~\eqref{eq:mass}, it yields
\be
\left|\dfrac{dM}{dt_{*}}\right|=2\sqrt{5.6\times 8\pi MH(t_{e})}\dfrac{a(t_{e})}{a(t_{*})}M_{\mrm{p}}.
\ee
Then $f(M)$ can be expressed in terms of the nucleation time
\be
f(M)=
\dfrac{M^{3 / 2} e^{-S_{E}\left(t_{*}\right)} a^{4}\left(t_{*}\right) H^{4}\left(t_{e}\right)}{2 \rho_{DM} a(t_{e}) a^{3}\left(t_{0}\right)M_{\mrm{p}}}\sqrt{\dfrac{1}{5.6\times 8\pi H(t_{e})}}.
\label{eq:fM}
\ee

The mass function of PBHs is plotted in Fig.~\ref{fig:mass_func} for the parameters sets listed in Table~\ref{ta:para}.
For the parameter set 1, $f(M)$ peaks at $M\sim 10^{17}\mrm{g}$.
For the parameter set 2, $f(M)$ peaks at $M\sim 10^{20}\mrm{g}$ and the PBHs could constitute all the dark matter.
For the parameter set 3, $f(M)$ peaks at $M\sim 10^{34}\mrm{g}$ to explain binary black hole merger rate detected by LIGO.

Due to the exponential dependence of $S_{E}$ in Eq.~\eqref{eq:S_E} and Eq.~\eqref{eq:lambda}, most DWs nucleate at the time $\phi=\phi_{c}$. In principle this model can provide DW radius concentrated upon any scale of cosmological interest.

The results hold for any inflationary potentials which have the same behavior near $\phi=\phi_{c}$. If the $ S_{E} $ reaches the minimum and then increases, the mass function will have a spike-like structure, independent of the dynamics of the scalar fields when $ t\neq t_{*} $.

The upper bound of PBH mass is given by the case where DWs are formed at the beginning of inflation. The maximal  radius at the end of inflation and the PBH mass are, respectively
\begin{eqnarray}
&&R_{max}=\dfrac{e^{N}}{H(t_{i})},\nonumber\\
&&M_{max}=M=5.6 \times 8 \pi (e^{N}/H(t_{i}))^{2} H\left(t_{e}\right) M_{\mathrm{p}}^{2},
\end{eqnarray}
where $t_{i}$ denotes the time when the inflation starts. The maximal mass is independent of $\sigma$ and only depends on the dynamics of the inflaton. Setting $N=50$ and $H(t_{i})=H(t_{e})=10^{-8}M_{\mrm{p}}$, $M_{max}$ is estimated to be $4\times 10^{48}\mrm{g}$.


\section{Conclusion and Discussion}
\label{sec:con}
In this paper, we have investigated the formation of PBHs from DWs in a two-field inflationary model.
DWs nucleate during inflation through quantum tunneling.
Since the Euclidean action $ S_{E} $ of DWs changes with time,
DWs nucleate mainly at the time when $ S_{E} $ reaches the minimum.
Since the PBHs mass depends on the nucleation time,
the mass range in general is narrow. The predicted mass function of PBHs has a spike-like structure. PBHs with the mass centered around $ 10^{20}\mrm{g} $ may constitute all dark matter. Since DWs nucleated through quantum tunneling are spherically symmetric, according to Birkhoff's theorem, the spherical DWs collapse cannot emit gravitational waves, in contrary to other mechanisms for the PBH formation,
in which measurements of the stochastic gravitational wave background put constraints on the abundance of PBHs with mass around $ 10^{20}\mrm{g} $ by LISA~\cite{Audley:2017drz} and Taiji~\cite{Guo:2018npi}.

PBHs can inversely give constraint on $ S_{E} $ during inflation. A conserved estimation indicates that $ S_{E}>1 $ for the PBH mass larger than $ 10^{15}\mrm{g} $. Taking $ H\sim 10^{-8}M_{\mrm{p}} $, it requires $ \sigma>5\times10^{22}M_{\mrm{p}} $. For PBHs with mass less than $ 10^{15}\mrm{g} $, in principle there is no lower bound on $ S_{E} $. The Universe might undergoes a period of the PBH-dominated era. 
The Hawking radiation of the overproduced PBHs can reheat the Universe~\cite{Hidalgo:2011fj}.  The coupling between the inflaton and other fields can be small without affecting the reheating process. The inflaton which does not form PBHs may serve as dark matter.

The nucleation rate is extremely sensitive to the Euclidean action of DWs in our model so that the fine-tuning problem still remains unsolved. DWs can also form during preheating after inflation if the potential has more than one degenerate vacua.
The radius and lifetime of DWs are sensitive to the potential barrier. Such DWs which stretch outside the Hubble horizon will result in massive PBHs. This topic is left to future investigation.


\begin{acknowledgements}
	We are very grateful to Heling Deng and Chengjie Fu for fruitful discussions. This work is supported in part by the National Natural Science Foundation of China Grants
	No.11690021, No.11690022, No.11575272, No.11851302, No.11435006 and No.11821505,
	in part by the Strategic Priority Research Program of the Chinese Academy of Sciences Grant No. XDB23030100
	and by Key Research Program of Frontier Sciences, CAS.
\end{acknowledgements}

\end{document}